\documentclass[aps, prl, twocolumn, superscriptaddress, notitlepage,floatfix]{revtex4-2}
\usepackage{float}
\usepackage{bbm}
\usepackage{epsfig}
\usepackage{epstopdf}
\usepackage{graphicx}
\usepackage{amsmath,amssymb}
\usepackage{amsmath,bm}
\usepackage{amsmath,esint}
\usepackage{physics}
\usepackage{color}
\usepackage{hyperref}
\usepackage{url}
\usepackage{lineno,blindtext}
\usepackage[caption=false]{subfig}
\usepackage{siunitx, booktabs}
\usepackage{diagbox, eqparbox, hhline}
\usepackage{soul}
\usepackage{xcolor}
\usepackage[normalem]{ulem}
\newcolumntype{P}[1]{>{\centering\arraybackslash}p{#1}}

\begin{document}

\title{Giant Response and Harmonic Generation in Néel-Torque Antiferromagnetic Resonance}

\author{Kuangyin Deng}
\email{kuangyid@ucr.edu}
\affiliation{Department of Electrical and Computer Engineering,  University of California, Riverside, California 92521, USA}
\author{Ran Cheng}
\email{rancheng@ucr.edu}
\affiliation{Department of Electrical and Computer Engineering,  University of California, Riverside, California 92521, USA}
\affiliation{Department of Physics and Astronomy, University of California, Riverside, California 92521, USA}
\affiliation{Department of Materials Science and Engineering, University of California, Riverside, California 92521, USA}

\begin{abstract}
We theoretically investigate the resonant and higher order magnetic responses of a collinear antiferromagnet induced by Néel spin-orbit torques (NSOTs). By deriving the dynamical susceptibilities up to the third harmonic, we find remarkable NSOT-induced amplifications of the linear and nonlinear magnetic dynamics by orders of magnitude compared to conventional spin-orbit torques, enabling highly-efficient frequency conversion in the terahertz frequency range. From the effective dynamics, we uncover the strong coupling between the Néel vector and the driving field of NSOTs, providing a physical explanation of the gigantic responses at all orders. We then propose a multilayer antiferromagnetic nano-device leveraging the gigantic harmonic generation to achieve unprecedented frequency amplifiers and converters. Our work uncovers a previously overlooked role of the NSOTs in nonlinear dynamics.
\end{abstract}

\maketitle 

\textit{Introduction.---} Antiferromagnetic materials (AFMs) are promising candidates for next-generation spintronic devices thanks to their terahertz (THz) dynamics and robustness against magnetic perturbations—properties that not only enable ultrafast operations but also mitigate cross-talks in high-density architectures~\cite{jungwirth2016antiferromagnetic,BaltzRMP2018,vsmejkal2018topological,gomonay2014spintronics,fukami2016magnetization,kampfrath2011coherent}. However, across the THz range ($10^{11}-10^{13}$ Hz), an essential obstacle is the scarcity of compact, efficient, room-temperature sources—--the long-noted “terahertz gap”. Rather than relying on native THz emitters, a pragmatic route is to \emph{up-convert and amplify} mature lower-frequency seeds; high-efficiency frequency multipliers thereby expand accessible THz channels while reducing reliance on cryogenic photonic or high-order electronic sources~\cite{Tonouchi2007NP,Nagatsuma2016NatPhotonics,Dhillon2017Roadmap}. Although AFMs could, in principle, act as standalone emitters, tunable on-chip operation is hampered by threshold/heating constraints and limited agility~\cite{demidov2012magnetic,khymyn2018ultra,chumak2015magnon}. This motivates employing AFMs as frequency amplifiers, leveraging their ultrafast bandwidth to translate modest seeds into THz carriers and thereby enlarge the usable channel set.

Nevertheless, it remains elusive as to the nonlinear magnetic dynamics in AFMs, especially the higher harmonic generation (HHG) controllable through current-induced torques. In the study of ferromagnetic resonance (FMR)~\cite{kittel1948theory}, HHG provides critical insights into nonlinear damping and magnon interactions, with applications ranging from frequency multipliers to spin-wave amplifiers~\cite{stancil2009spin,zhou2024skin}. However, exciting higher harmonics in AFMs presents fundamental challenges:
(1) The Néel vector, as a staggered order parameter, does not couple directly to a uniform magnetic field or conventional spin torques that act on the net magnetization. Consequently, the Néel vector dynamics is typically driven \textit{indirectly} and \textit{weakly} by external stimuli, hence requiring substantial power densities and risking excessive Joule heating~\cite{gurevich2020magnetization};
(2) Strong drives can induce parametric instabilities, such as Suhl processes~\cite{gurevich2020magnetization, suhl1957theory}, in which the primary mode ($\bm{k}=0$) decays into pairs of magnons (with $\bm{k}=\pm \bm{k}_0$). While such decay channels are forbidden in the linear response regime, an intense driving field can activate magnon-magnon interactions, thereby opening nonlinear scattering channels. These instabilities redistribute energy of the driven mode among a spectral continuum, leading to significant resonance broadening that obscures the higher-harmonic peaks. (3) Conventional approaches to realizing nonlinear responses usually rely on extrinsic mechanisms such as structural engineering~\cite{sun2019giant,song2022higher}, energy-barrier (\textit{i.e.}, magnetic anisotropy) engineering~\cite{demidov2012magnetic,khymyn2018ultra}, nonlinear spin current injection~\cite{omar2020nonlinear}, and spin-phonon coupling~\cite{yarmohammadi2024tera}, all of which suffer from certain physical limitations and inevitably complicate integration into scalable AFM-based devices.

Néel spin–orbit torques (NSOTs), or simply Néel torques~\cite{vzelezny2014relativistic,wadley2016electrical,gomonay2016high,vzelezny2017spin,bodnar2018writing,chen2019electric,manchon2019current,shao2023neel,behovits2023terahertz,feng2025intrinsic}, which typically require the inversion symmetry breaking in AFMs, offer a promising opportunity to enable \textit{direct} and \textit{strong} coupling between electric currents and the Néel vector dynamics without relying on the small magnetization. This is because the NSOT arises from the staggered non-equilibrium spins (\textit{i.e.}, opposite spin polarizations acting on opposite sublattices) that intrinsically matches the symmetry of Néel vector, unleashing a unique mechanism to leverage the THz dynamics of AFMs. A recent study reported unexpectedly strong linear responses of the Néel vector to the NSOTs~\cite{tang2025neel}, which leads to an amplification of the linear dynamical susceptibility by over two orders of magnitude compared to conventional spin torques.

Nevertheless, a crucial problem remains virtually unexplored: can NSOTs, beyond enabling linear control of the Néel order, also serve as an efficient driver of nonlinear spin dynamics and why? In particular, it is tempting to ask whether NSOTs can excite the HHG of Néel vector dynamics without resorting to any extrinsic mechanism such as symmetry-breaking magnetic anisotropy.

In this Letter, we develop a theoretical framework for the responses at all orders and the HHG in collinear AFM systems driven by NSOTs, offering a direct pathway to access the versatile nonlinear spin dynamics that \textit{does not} rely on auxiliary mechanisms. We calculate the dynamical susceptibility for each harmonic response and benchmark the result against that of conventional spin-orbit torque (SOT)~\cite{cheng2014spinpumping}, revealing orders-of-magnitude enhancement in the first, second, and third harmonics. The underlying physical reason is that, under NSOTs, the driving field couples strongly to the N\'eel vector through the exchange field in its effective dynamics (shown in the \textbf{End Matter}). As a demonstration, we propose a scalable AFM multilayer heterostructure that can be utilized as a highly efficient frequency converter in the THz regime with very low driving power, which provides concrete guidance for designing frequency multipliers in magnetic systems.

\textit{Model and Nonlinear Harmonic Response.---} To quantify the linear and nonlinear dynamics of an AFM to the NSOT, we calculate the dynamical susceptibility for each harmonic response. We consider a collinear AFM with an easy axis along $\hat{z}$, whose sublattice magnetic moments are described by the unit vectors $\bm{m}^{A(B)}=\bm{M}^{A(B)}/M$ with $M=|\bm{M}^{A(B)}|$. These vectors obey the Landau–Lifshitz–Gilbert (LLG) equations:
\begin{align}
\dot{\bm{m}}^{A(B)} = -\gamma \bm{m}^{A(B)} \cross \bm{H}_{\mathrm{eff}}^{A(B)} + \alpha \bm{m}^{A(B)} \cross \dot{\bm{m}}^{A(B)},\label{eq:LLGwhole}
\end{align}
where $\gamma>0$ is the gyromagnetic ratio, and $\alpha>0$ is the Gilbert damping constant, and $\dot{\bm{m}}^{A(B)}$ denotes the time derivative of $\bm{m}^{A(B)}$. The effective field for each sublattice reads
\begin{align}
\bm{H}_{\mathrm{eff}}^{A(B)}=H_0\hat{z}-H_J\bm{m}^{B(A)}+H_{\parallel}m_z^{A(B)}\hat{z}+\bm{h}^{A(B)},
\label{eq:heff-total}
\end{align}
where $H_0\hat{z}$ is an external static magnetic field along the $z$-axis, $H_J>0$ is the exchange field, $H_{\parallel}>0$ is the anisotropy field, and $\bm{h}^{A(B)}$ is an AC driving field representing the spin torque (either NSOT or SOT). In equilibrium, $\bm{m}^{A(B)}_0=(0,\,0,\,\pm 1)$, and we let
\begin{align}
\bm{m}^{A(B)}=\bm{m}^{A(B)}_0+\delta\bm{m}^{A(B)}, \label{eq:mtotal}
\end{align}
with $\delta\bm m^{A(B)}$ being small deviations.

\begin{figure}[t]
\centering
\includegraphics[width=1\linewidth]{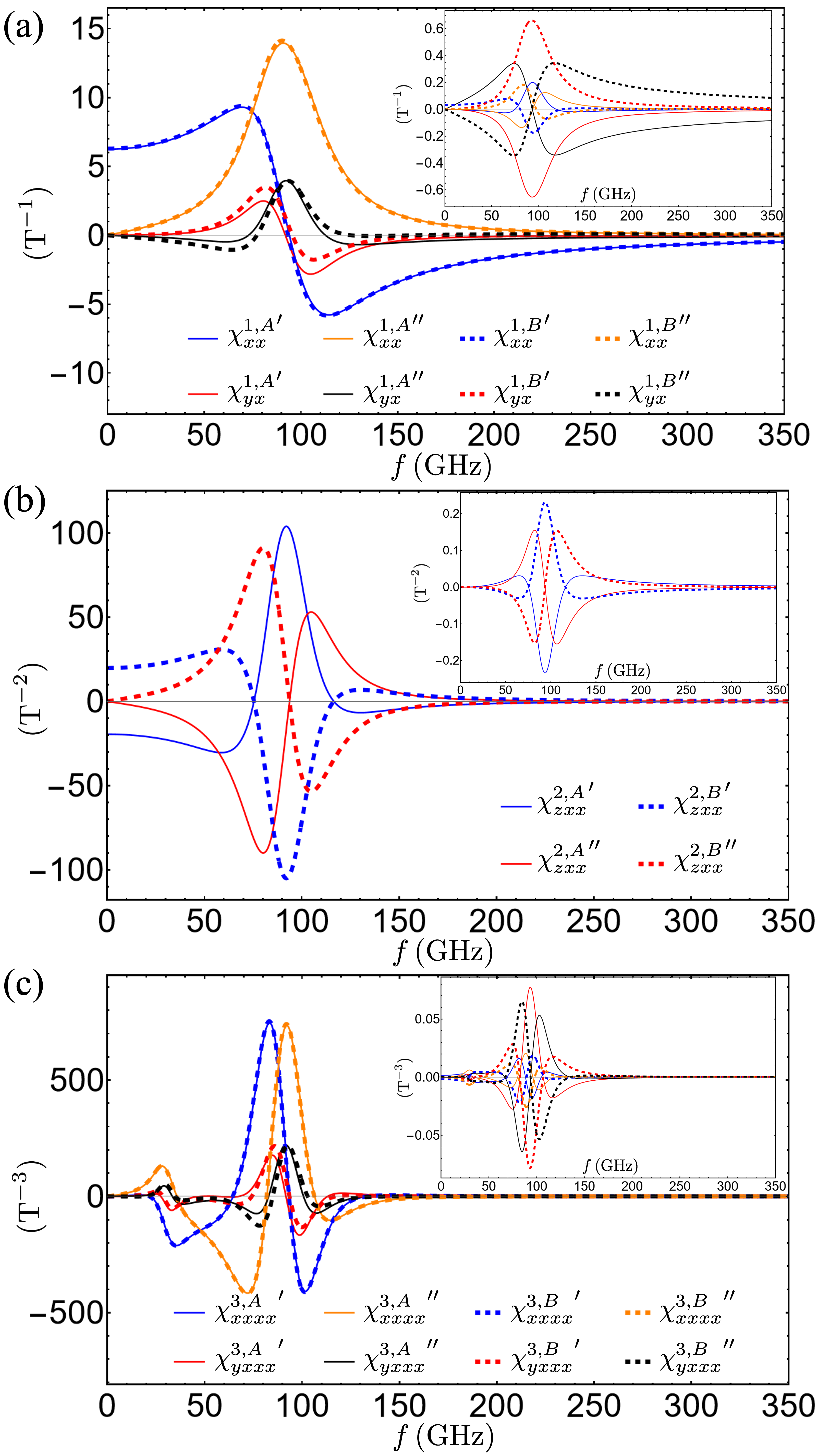}
\caption{Real ($\chi'$) and imaginary parts ($\chi''$) of (a) the first‐, (b) second‐, and (c) third‐order susceptibilities under the NSOT.  Solid (dashed) lines denote sublattice A (B); their real and imaginary components are distinguished by colors.  Insets reproduce the same susceptibilities under conventional SOT using identical line styles and colors.  All harmonic responses are markedly enhanced by the NSOT compared to SOT. Parameters: $\gamma=2\pi\cdot 28.02$ GHz$\cdot$T$^{-1}$, $\alpha=0.02$, $H_0=0.2$ T, $H_J=35$ T, $H_\parallel=0.16$ T, and $f=\omega/2\pi$.}
\label{fig:chiAandB}
\end{figure}

Under conventional SOTs, the driving field can simply be an applied AC magnetic field so that $\bm{h}^A=\bm{h}^B=\bm{h}$. By contrast, in the case of NSOTs, the driving field is generated via intrinsic mechanisms~\cite{vzelezny2014relativistic,feng2025intrinsic,tang2025neel}, satisfying the staggered relation $\bm{h}^A=-\bm{h}^B=\bm{h}$. A prominent class of AFM systems capable of generating such a staggered field are those with $\mathcal{PT}$ symmetry, where the symmetry enforces opposite spin polarizations on the two magnetic sublattices under a time-dependent electric field. This symmetry-based mechanism includes both the Berry-curvature–mediated generation of NSOT via the time derivative of the electric field~\cite{feng2025intrinsic} and the direct generation of NSOT under a time-dependent electric field~\cite{tang2025neel}. For instance, in a $\mathcal{PT}$-symmetric AFM with strong spin-orbit coupling, an applied electric field can generate the adiabatic currents of topological electrons, which in turn exert such a staggered field on each sublattice~\cite{tang2025neel}. To capture effects beyond linear response, we expand $\delta\bm{m}^{A(B)}$ in Eq.~\eqref{eq:mtotal} into a series of harmonics~\cite{topping2018ac}:
\begin{align}
\delta m_i^{A(B)}=&\chi^{1,A(B)}_{ij}h_j^{A(B)} + \chi^{2,A(B)}_{ijk}h_j^{A(B)}h_k^{A(B)} \nonumber\\
&\quad+ \chi^{3,A(B)}_{ijkl}h_j^{A(B)}h_k^{A(B)}h_l^{A(B)} +\cdots,
\label{eq:dmgeneral}
\end{align}
where $\chi^{\sigma,A(B)}$ denotes the $\sigma$th-order susceptibility of sublattice $A (B)$. For simplicity, we consider a NSOT field polarized along the $x$ direction such that $\bm{h}^A=-\bm{h}^B=he^{-\mathrm{i}\omega t}\hat{x}$, which leads to~\footnote{In the conventional-SOT case, there is no sign difference for A and B.}
\begin{align}
\delta m_i^{A(B)}=&\pm\chi^{1,A(B)}_{ix}he^{-\mathrm{i}\omega t}+\chi^{2,A(B)}_{ixx}h^2e^{-2\mathrm{i}\omega t}\nonumber\\
&\pm\chi^{3,A(B)}_{ixxx}h^3e^{-3\mathrm{i}\omega t} +\cdots
\label{eq:dmi}
\end{align}
where ``$+$" and ``$-$" are for sublattice A and B, respectively. Substituting Eq.\eqref{eq:dmi} and its time derivative into Eq.\eqref{eq:mtotal}, and then inserting the result into Eq.~\eqref{eq:LLGwhole}, we can collect and compare terms at each harmonic $e^{-\mathrm{i}\omega t}$, $e^{-2\mathrm{i}\omega t}$, and $e^{-3\mathrm{i}\omega t}$. Since Eq.~\eqref{eq:LLGwhole} involves two vector equations (for A and B sublattices) with each including three components, matching the coefficients for all the three harmonic orders produces a total of $18$ scalar equations, which exactly correspond to the $18$ unknown susceptibility components $\chi_{ix}^{1,A(B)}$, $\chi_{ixx}^{2,A(B)}$, and $\chi_{ixxx}^{3,A(B)}$. Because higher‐order susceptibilities depend on lower‐order ones, we have to solve them sequentially: first determine all linear susceptibilities, then the second‐order components, followed by the third‐order ones. Following this procedure, we find the following components vanish identically:
\begin{align}
\chi^{1,A(B)}_{zx}=\chi^{2,A(B)}_{xxx}=\chi^{2,A(B)}_{yxx}=\chi^{3,A(B)}_{zxxx}=0,
\end{align}
leaving $10$ non-zero complex variables to solve. The vanishing components above can be understood from symmetry analysis. With the easy axis along $z$, the equilibrium Hamiltonian has SO(2) rotational symmetry in the $xy$-plane. For an in-plane driving field $\bm{h}$ inducing $\delta\bm{m} = (\delta m_x,\,\delta m_y,\,\delta m_z)$, reversing the field to $-\bm{h}$ corresponds to a $\pi$-rotation about $\hat{z}$, yielding $\delta\bm{m}' = (-\delta m_x,\,-\delta m_y,\,\delta m_z)$. For both SOT and NSOT cases, this symmetry enforces that all even-order terms of $\delta m_{x(y)}$ and all odd-order terms of $\delta m_z$ vanish [see Eq.~\eqref{eq:dmgeneral}]. Although the driving breaks the symmetry, the susceptibilities—being equilibrium properties—reflect the symmetry prior to excitation.

Unfortunately, analytical expressions for these $10$ variables are too lengthy to present. So instead, we plot their real ($\chi'$) and imaginary ($\chi''$) parts~\footnote{If we instead adopt the alternative convention that $\bm{h}(t) = \bm{h}e^{\mathrm{i}\omega t}$, the imaginary part $\chi''$ flips sign.} in Fig.~\ref{fig:chiAandB} for empirical materials parameters $\gamma=2\pi\cdot 28.02$~GHz$\cdot$T$^{-1}$, $\alpha=0.02$, $H_0=0.2$~T, $H_J=35$~T, $H_\parallel=0.16$~T, and $f=\omega/2\pi$. While $\chi^{q,A(B)}$ is independent of the magnitude of the driving field $\abs{\bm{h}}$, the condition $\abs{\bm{h}} \ll \frac{\omega_0^2}{2\gamma^2 H_J} \approx 0.18~\text{T}$ should be generally satisfied ($\omega_0=2\pi f_0$ is the resonant frequency), ensuring that the field acts only as a weak perturbation around the equilibrium state, as estimated from the effective dynamical Eq.~\eqref{eq:nsotncrossddn} for the N\'eel vector in the \textbf{End Matter}. The main panels display the NSOT results for each harmonic, while the insets show the corresponding SOT responses under the same set of parameters. For each sublattice (A or B), the linear susceptibility (\textit{i.e.}, first harmonic) is amplified by roughly two orders of magnitude under NSOT as compared to SOT, whereas the second and third harmonics are enhanced by about three and four orders of magnitude, respectively.  Furthermore, the odd‐order responses swap their $x$–$y$ ordering: for example, in the vicinity of the resonance point, $\bigl|\chi^{1,A(B)}_{xx}\bigr|>\bigl|\chi^{1,A(B)}_{yx}\bigr|$ for the NSOT, while $\bigl|\chi^{1,A(B)}_{xx}\bigr|<\bigl|\chi^{1,A(B)}_{yx}\bigr|$ for the SOT. This asymmetric amplification of higher harmonics stems from the unique symmetry of nonlinear coupling in AFMs driven by NSOTs.

Because NSOTs introduce a staggered drive $\bm h^A = +\bm h$ and $\bm h^B = -\bm h$, one might think to redefine $\tilde\chi^B = -\chi^B$ (with $\tilde\chi^A = \chi^A$) so that both sublattices see an effective uniform field $\bm h$, reducing to the conventional‐SOT scenario.  However, this fails: the minus sign enters not only the driving term in $\delta\bm m^{A(B)}$ (Eq.~\eqref{eq:dmi}) but also the effective fields $\bm H_{\mathrm{eff}}^{A(B)}$ in Eq.~\eqref{eq:heff-total}.  As a result, every susceptibility component is modified—not just its sign—and the NSOT case cannot be mapped onto a uniform‐field SOT by a simple redefinition.

\textit{Power Absorption.---} Having established the dynamical susceptibility of each sublattice, we are ready to evaluate the power absorption rate under NSOT and that under conventional SOT.  The instantaneous absorbed power for a single magnetic moment under harmonic excitation is given by~\cite{slichter2013principles}:
\begin{align}
P^{A(B)}(t)=\frac{1}{2}\mathrm{Re}\left[\frac{d\bm{m}^{A(B)}(t)}{dt}\cdot{\bm{h}^{A(B)}}^*(t)\right],\label{eq:pab}
\end{align}
and the total instantaneous power absorption is $P^{\mathrm{tot}}(t)=P^{A}(t)+P^{B}(t)$. In the NSOT case, we have $\bm{h}^A =-\bm{h}^B = \bm{h}$, so the total power becomes
\begin{align}
P^{\mathrm{tot}}(t)=\frac{1}{2}\mathrm{Re}\left[\frac{d\bm{n}(t)}{dt}\cdot \bm{h}^*(t)\right],\label{eq:ptotalwithneel}
\end{align}
where $\bm{n}\equiv (\bm{m}^A-\bm{m}^B)/2$ is the Néel vector. By sharp contrast, the total power associated with the SOT is determined by the small magnetization $\bm{m}\equiv (\bm{m}^A+\bm{m}^B)/2$ in the form of $\mathrm{Re}[\dot{\bm{m}}(t)\cdot\bm{h}^*(t)]/2$. Inserting the harmonic expansions of Eq.~\eqref{eq:dmgeneral} into Eq.~\eqref{eq:ptotalwithneel} yields
\begin{align}
P^{\mathrm{tot}}
&=\frac{1}{2}\mathrm{Re}\left[(-\mathrm{i}\omega) h_i^*(\chi^{1,A}_{ij}+\chi^{1,B}_{ij})h_j\right.\nonumber\\
&\quad+(-2\mathrm{i}\omega) h_i^*(\chi^{2,A}_{ijk}-\chi^{2,B}_{ijk})h_j h_k\nonumber\\
&\quad\left.+(-3\mathrm{i}\omega) h_i^*(\chi^{3,A}_{ijkl}+\chi^{3,B}_{ijkl})h_j h_k h_l\right],
\end{align}
from which one can define the total susceptibility at each harmonic order:
\begin{align}
\chi^{q,\mathrm{tot}}=&\chi^{q,A}+(-1)^{q-1}\chi^{q,B},
\quad q = 1,2,3,\dots
\label{eq:totalchi-nsot}
\end{align}
This alternating “even–odd” pattern originates from the staggered nature of the NSOT: odd-order harmonics add the sublattice contributions, whereas even-order harmonics enter with opposite signs. This interference introduces an extra enhancement besides the large single-sublattice amplification shown in Fig.~\ref{fig:chiAandB}, rendering the susceptibilities associated with the two sublattices always adding up constructively. Note that $\bm{h}$ may represent an effective field proportional to, for example, the time derivative of an electric field~\cite{feng2025intrinsic}. By contrast, under conventional SOT where $\bm h^A = \bm h^B = \bm h$, similar steps yield
\begin{align}
\chi^{q,\mathrm{tot}}=&\chi^{q,A}+\chi^{q,B},
\quad q = 1,2,3,\dots
\label{eq:totalchi-sot}
\end{align}
which, when considering the built-in opposite sign between A and B shown in Fig.~\ref{fig:chiAandB}, leads to overall destructive contributions from the two sublattices.

\begin{figure}[t]
\centering
\includegraphics[width=1\linewidth]{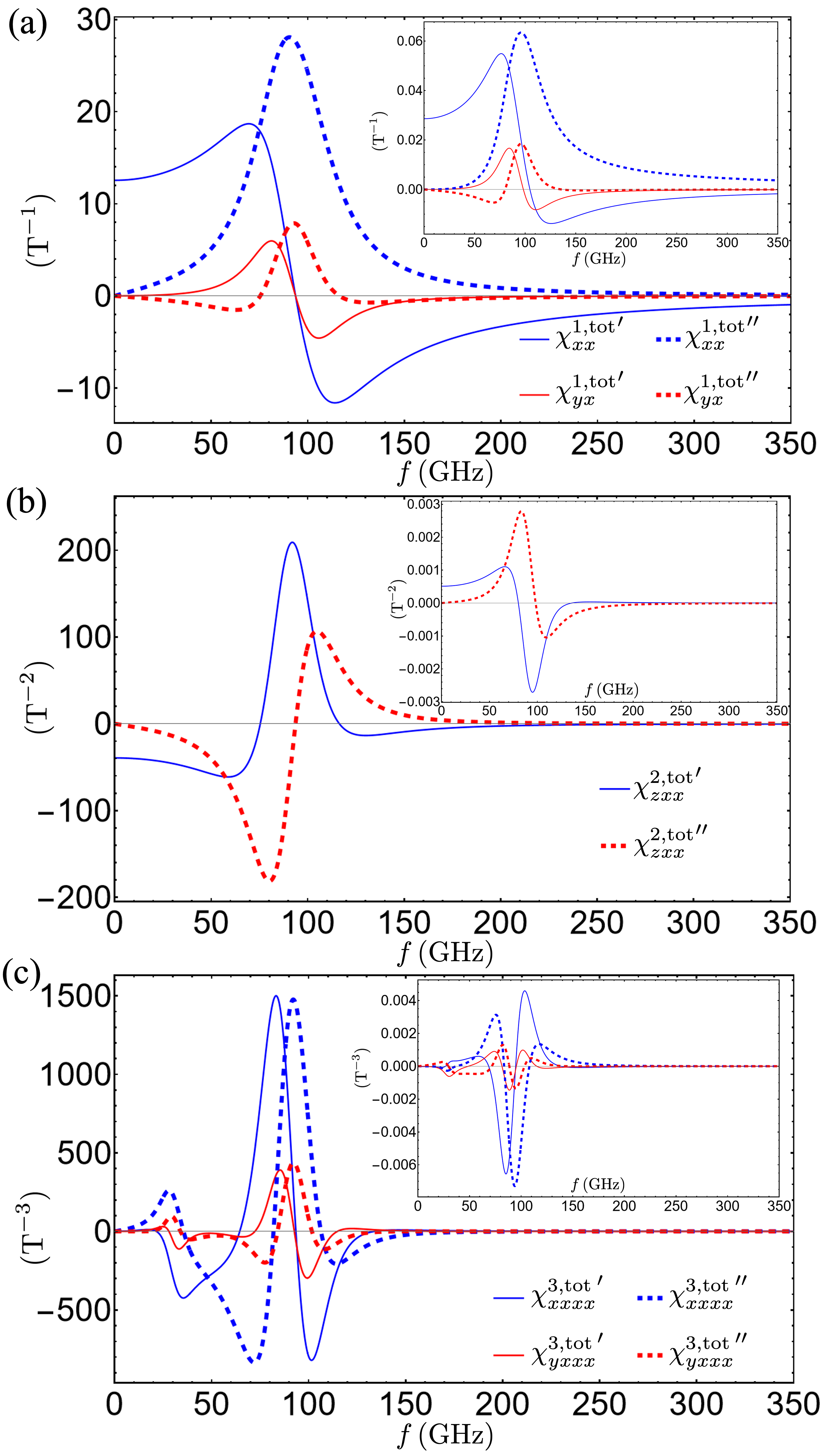}
\caption{Real (${\chi^{\mathrm{tot}}}'$, solid) and imaginary (${\chi^{\mathrm{tot}}}''$, dashed) parts of the total susceptibilities at their (a) first‐order, (b) second‐order, and (c) third‐order under the NSOT.  Insets reproduce the same plots for conventional SOT with identical line styles and colors.  All harmonic responses are markedly amplified by the NSOT compared to SOT.  Parameters are the same as in Fig.~\ref{fig:chiAandB}.}
\label{fig:chitotal}
\end{figure}

Using the same set of parameters, Fig.~\ref{fig:chitotal} shows the real (${\chi^{q,\mathrm{tot}}}'$) and imaginary (${\chi^{q,\mathrm{tot}}}''$) components of the total susceptibilities. The main panels (insets) display the NSOT (SOT) results.  Compared to their counterparts for individual sublattices, the total susceptibilities under the NSOT are indeed amplified even more dramatically—approximately three, five, and six orders of magnitude for the first, second, and third harmonics, respectively. This fully agrees with our expectations on Eqs.~\eqref{eq:totalchi-nsot} and~\eqref{eq:totalchi-sot}. Although, in our example, the resonant frequency is about $0.1$ THz, our theory applies to all collinear AFMs, thereby spanning a wide range of frequency channels in the THz region.

We now provide an intuitive physical explanation to the giant amplification. By eliminating the small magnetic moment $\bm{m}$, the effective dynamics of the N\'eel vector is (see the \textbf{End Matter}),
\begin{align}
\bm{n}\cross \ddot{\bm{n}}=&-\gamma H_0 (\dot{n}_z\bm{n}-n_z\dot{\bm{n}}) -\gamma(2\alpha H_J+H_0)\bm{n}\cross \dot{\bm{n}}\nonumber\\
&+\gamma^2 (2H_J H_{\parallel}-H_0^2)n_z\bm{n}\cross\hat{z}\nonumber\\
&+2\gamma^2 H_J\bm{n}\cross\bm{h},\label{eq:neeldynamics}
\end{align}
where the last term proportional to $\gamma H_J$ indicates a rather strong coupling of $\bm{n}$ to the driving field $\gamma \bm{h}$ (in the frequency domain). By contrast, a conventional SOT couples to $\bm{n}$ through the much weaker Zeeman interaction $\gamma H_{0}$ and the driving frequency $\omega$ [see Eq.~(A10) in the \textbf{End Matter}]. In real materials, $\gamma H_J$ typically exceeds $\gamma H_{0}$ and $\omega\sim\omega_0$ near resonance by several orders of magnitude (see the caption of Fig.~\ref{fig:chiAandB}), rendering a remarkably stronger driving efficiency for the NSOT as compared to the SOT. According to Eq.~\eqref{eq:ptotalwithneel}, the total absorbed power $P^{\mathrm{tot}}$ is proportional to $\dot{\bm{n}}$, so the boosted coupling strength shown above is inherited at all harmonics, including both linear and nonlinear responses.

The average power absorption is defined as $\bar{P}^{\mathrm{tot}}(\omega)=\frac{1}{T}\int_0^T P^{\mathrm{tot}}(t)dt$, where $T$ is the period of oscillation of the driving field. With $\bm{h}=he^{-\mathrm{i}\omega t}\hat{x}$, this evaluates to
\begin{align}
\bar{P}^{\mathrm{tot}}(\omega)=&\frac{1}{2}\omega {\chi^{1,\mathrm{tot}}_{xx}}'' h^2,
\end{align}
which holds for both NSOT and SOT. Because the higher harmonics ($e^{-\mathrm{i}n\omega t}$, $n\neq 1$) are orthogonal to $e^{-\mathrm{i}\omega t}$ within $[0,T]$, they do not contribute to the average total power.  

Although the net average power absorption at higher harmonics vanishes under single-frequency excitation of our AFM, these harmonics can excite a secondary resonator coupled to the AFM, thereby transferring energy into that system. Consequently, an AFM driven by NSOT not only generates large, amplified signals at $2\omega$ and $3\omega$, but also efficiently channels power into any suitably tuned device, acting as an intrinsic frequency converter and emitter, which is crucial for high-frequency spintronic applications.

\begin{figure}[t]
\centering
\includegraphics[width=0.85\linewidth]{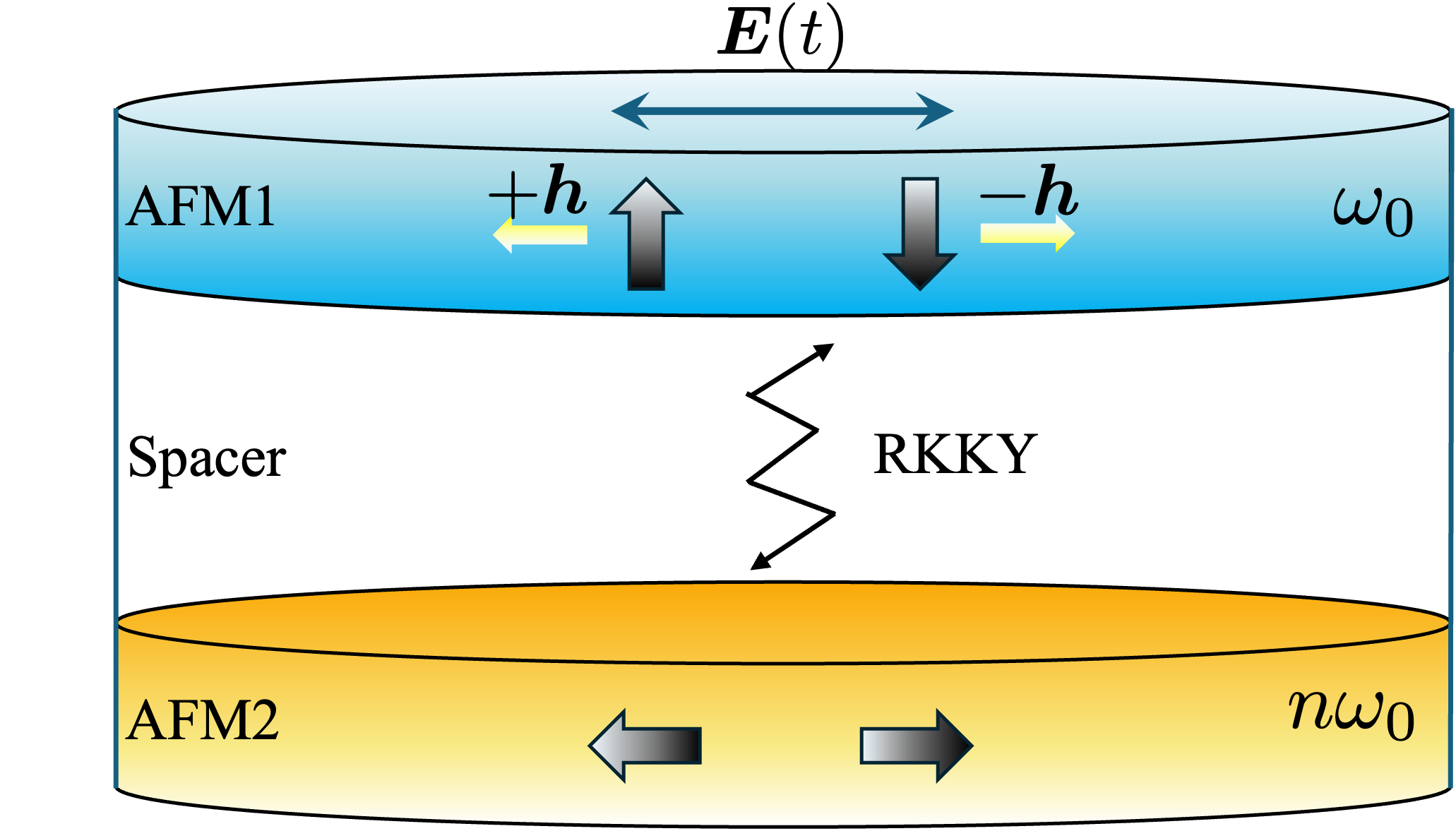}
\caption{Schematic of the proposed AFM frequency converter.  A $\mathcal{PT}$‐symmetric AFM1 with eigenfrequency $\omega_0$ is driven by NSOTs, generating higher harmonics.  Through the RKKY coupling, the $n$th harmonic of AFM1 resonantly excites a second AFM layer (AFM2) whose resonance frequency is $n\omega_0$, enabling energy transfer from AFM1 to AFM2, effectively converting a frequency source of $\omega_0$ into $n\omega_0$.}
\label{fig:freq-trans}
\end{figure}

\textit{Proposed Device.---} Building on our theoretical demonstration of strong HHG in NSOT-driven AFMs, we propose a nanoscale frequency converter comprising of two AFM layers separated by a thin nonmagnetic spacer as illustrated in Fig.~\ref{fig:freq-trans}. The first layer (AFM1, $\mathcal{PT}$-symmetric) is driven at its resonance frequency $\omega_0$ by the NSOT induced by a time‐dependent electric field, generating enhanced second- and third-harmonic signals. These harmonics then couple to a different layer (AFM2) through the RKKY exchange interaction, given that AFM2’s resonance frequency matches $n\omega_0$ for $n=2$ or $3$. In this coupled system, AFM1’s time‐averaged power includes the power absorption from the NSOT and that delivered to AFM2, where the latter results in negative work on AFM1 due to back-actions. By contrast, existing magnonic converters depend on engineered microstructures or hybrid couplings (e.g., spin-phonon interactions) and lack intrinsic, symmetry-allowed nonlinear mechanisms~\cite{chumak2015magnon}. Our design instead harnesses the giant intrinsic higher-harmonic responses, enabling substantial HHG with only modest driving power and can avoid parametric instabilities, thereby achieving efficient, intrinsic THz frequency conversion in AFM heterostructures.

Beyond the bilayer device, we propose a crossbar‐array architecture in which orthogonal rows and columns of the above AFM bilayer units form a matrix of harmonic converters. Each bilayer unit, driven by NSOTs, transforms an input signal near its own resonant frequency into higher harmonics and directs the selected harmonic along the intersecting channel. By cascading multiple bilayers in series, successive stages of frequency multiplication and signal boosting can be realized. This grid architecture supports parallel frequency channels for high‐density integration and enables multiplexed spintronic communication with minimal crosstalk~\cite{zhou2022orthogonal,qi2024antiferromagnetic,locatelli2014spin,grollier2020neuromorphic}.

\begin{acknowledgements}
This work was supported by the National Science Foundation under Award No. DMR-2339315.
\end{acknowledgements}

%

\clearpage
\appendix
\onecolumngrid
\begin{center}
  \textbf{\large End Matter}\label{app-A}
\end{center}
\twocolumngrid

\setcounter{equation}{0}                       
\renewcommand{\theequation}{A\arabic{equation}}

\textit{Appendix A:} Here we show the effective dynamics of the Néel vector of our system under a field-like SOT and NSOT, respectively. This provides a physical understanding of the amplification of the response functions under NSOTs.

We define the total magnetic moment $\bm{m}=(\bm{m}^A+\bm{m}^B)/2$ and the Néel vector $\bm{n}=(\bm{m}^A-\bm{m}^B)/2$. Eq.~\eqref{eq:LLGwhole} can be re-written as
\begin{align}
\dot{\bm{m}}+\dot{\bm{n}}=& -\gamma (\bm{m}+\bm{n})\cross \left[H_0\hat{z}-H_J(\bm{m}-\bm{n})\right.\nonumber\\
&\left.+H_{\parallel}(m_z+n_z)\hat{z}+\bm{h}^A\right]\nonumber\\
&+\alpha (\bm{m}+\bm{n})\cross (\dot{\bm{m}}+\dot{\bm{n}}),\\
\dot{\bm{m}}-\dot{\bm{n}}=& -\gamma (\bm{m}-\bm{n})\cross \left[H_0\hat{z}-H_J(\bm{m}+\bm{n})\right.\nonumber\\
&\left.+H_{\parallel}(m_z-n_z)\hat{z}+\bm{h}^B\right]\nonumber\\
&+\alpha (\bm{m}-\bm{n})\cross (\dot{\bm{m}}-\dot{\bm{n}}).
\end{align}
To capture the effective dynamics, we only need to keep the leading order terms. Considering the AFM order with $\bm{m}\cdot\bm{n}\approx 0$, $\abs{\bm{n}}\approx 1\gg \abs{\bm{m}}$ under small in-plane driving field $\abs{\bm{h}^{A(B)}}\ll H_0$, we have
\begin{align}
\dot{\bm{m}}+\dot{\bm{n}}=& -\gamma H_0(\bm{m}+\bm{n})\cross \hat{z}-2\gamma H_J \bm{m}\cross \bm{n}\nonumber\\
&-\gamma H_{\parallel}n_z\bm{n}\cross\hat{z}-\gamma \bm{n}\cross\bm{h}^A\nonumber\\
&+\alpha (\bm{m}+\bm{n})\cross (\dot{\bm{m}}+\dot{\bm{n}}),\\
\dot{\bm{m}}-\dot{\bm{n}}=& -\gamma H_0(\bm{m}-\bm{n})\cross \hat{z}+2\gamma H_J \bm{m}\cross \bm{n}\nonumber\\
&-\gamma H_{\parallel}n_z\bm{n}\cross\hat{z}+\gamma \bm{n}\cross\bm{h}^B\nonumber\\
&+\alpha (\bm{m}-\bm{n})\cross (\dot{\bm{m}}-\dot{\bm{n}}).
\end{align}
Adding and subtracting them lead to
\begin{align}
\dot{\bm{m}}=& -\gamma H_0\bm{m}\cross \hat{z}-\gamma H_{\parallel}n_z\bm{n}\cross\hat{z} +\alpha \bm{n}\cross \dot{\bm{n}}\nonumber\\
&-\frac{1}{2}\gamma \bm{n}\cross\left(\bm{h}^A-\bm{h}^B\right),\label{eq:mdot-raw}\\
\dot{\bm{n}}=& -\gamma H_0\bm{n}\cross \hat{z}-2\gamma H_J \bm{m}\cross \bm{n}\nonumber\\
&-\frac{1}{2}\gamma \bm{n}\cross\left(\bm{h}^A+\bm{h}^B\right).\label{eq:ndot-raw}
\end{align}

Now we need to eliminate the $\bm{m}$-dependence from Eq.~\eqref{eq:ndot-raw}. Applying the cross product $\bm{n}\cross$ to this equation and taking the leading order, we have
\begin{align}
\bm{n}\cross \dot{\bm{n}}=&-\gamma H_0 (n_z\bm{n}-\hat{z})-2\gamma H_J \bm{m} \nonumber\\
&+\frac{1}{2}\gamma \left(\bm{h}^A+\bm{h}^B\right),\label{eq:ncrossdotn}
\end{align}
where the term $\bm{n}\cdot\bm{h}^{A(B)}$ is neglected because $\bm{h}^{A(B)}$ is taken in-plane. This can be also expressed as
\begin{align}
2\gamma H_J \bm{m}=&-\bm{n}\cross \dot{\bm{n}}-\gamma H_0 (n_z\bm{n}-\hat{z})\nonumber\\
&+\frac{1}{2}\gamma \left(\bm{h}^A+\bm{h}^B\right).\label{eq:m-gamma}
\end{align}
Taking the time derivative of Eq.~\eqref{eq:ncrossdotn}, we obtain
\begin{align}
\bm{n}\cross \ddot{\bm{n}}=&-\gamma H_0 (\dot{n}_z\bm{n}-n_z\dot{\bm{n}})-2\gamma H_J \dot{\bm{m}} \nonumber\\
&+\frac{1}{2}\gamma \frac{d\left(\bm{h}^A+\bm{h}^B\right)}{dt}.\label{eq:ncrossddn}
\end{align}
Under a conventional SOT, the driving fields on the two sublattices are identical, \textit{i.e.}, $\bm{h}^A=\bm{h}^B=\bm{h}$. Utilizing Eqs.~\eqref{eq:mdot-raw} and~\eqref{eq:m-gamma}, Eq.~\eqref{eq:ncrossddn} becomes
\begin{align}
\bm{n}\cross \ddot{\bm{n}}=&-\gamma H_0 (\dot{n}_z\bm{n}-n_z\dot{\bm{n}})-2\gamma H_J \left(-\gamma H_0\bm{m}\cross \hat{z}\right.\nonumber\\
&\left.-\gamma H_{\parallel}n_z\bm{n}\cross\hat{z} +\alpha \bm{n}\cross \dot{\bm{n}}\right) +\gamma \dot{\bm{h}}\nonumber\\
=&-\gamma H_0 (\dot{n}_z\bm{n}-n_z\dot{\bm{n}}) -\gamma(2\alpha H_J+H_0)\bm{n}\cross \dot{\bm{n}}\nonumber\\
&+\gamma^2 (2H_J H_{\parallel}-H_0^2)n_z\bm{n}\cross\hat{z}\nonumber\\
&+\gamma^2 H_0\bm{h}\cross\hat{z} +\gamma \dot{\bm{h}},\label{eq:sotncrossddn}
\end{align}
where the driving function can be denoted as $\bm{F}_{\text{SOT}}=\gamma^2 H_0\bm{h}\cross\hat{z}+\gamma \dot{\bm{h}}=\gamma^2 H_0\bm{h}\cross\hat{z}-\mathrm{i}\gamma\omega\bm{h}$. Here $\omega$ is the driving frequency. In contrast, for NSOT the fields are staggered, $\bm{h}^A=-\bm{h}^B=\bm{h}$. Again, combining Eqs.~\eqref{eq:mdot-raw} and~\eqref{eq:m-gamma}, Eq.~\eqref{eq:ncrossddn} is reduced to
\begin{align}
\bm{n}\cross \ddot{\bm{n}}=&-\gamma H_0 (\dot{n}_z\bm{n}-n_z\dot{\bm{n}})-2\gamma H_J \left(-\gamma H_0\bm{m}\cross \hat{z}\right.\nonumber\\
&\left.-\gamma H_{\parallel}n_z\bm{n}\cross\hat{z} +\alpha \bm{n}\cross \dot{\bm{n}}-\gamma\bm{n}\cross\bm{h}\right) \nonumber\\
=&-\gamma H_0 (\dot{n}_z\bm{n}-n_z\dot{\bm{n}}) -\gamma(2\alpha H_J+H_0)\bm{n}\cross \dot{\bm{n}}\nonumber\\
&+\gamma^2 (2H_J H_{\parallel}-H_0^2)n_z\bm{n}\cross\hat{z}\nonumber\\
&+2\gamma^2 H_J\bm{n}\cross\bm{h},\label{eq:nsotncrossddn}
\end{align}
so that the driving function is $\bm{F}_{\text{NSOT}}=2\gamma^2 H_J\bm{n}\cross\bm{h}$. It is clear that Eq.~\eqref{eq:sotncrossddn} and~\eqref{eq:nsotncrossddn} are identical except the driving functions (last lines) $\bm{F}_{\text{SOT}}$ and $\bm{F}_{\text{NSOT}}$. For driving frequencies near resonance, $\omega\approx\omega_0$, a typical AFM satisfies (see Fig.~\ref{fig:chiAandB} and its caption for instance) $\gamma H_0<\omega\ll\gamma H_J$, such that $\abs{\bm{F}_{\text{SOT}}}\ll\abs{\bm{F}_{\text{NSOT}}}$. Hence, under regular SOT the N\'eel vector couples to the driving field only \textit{indirectly and weakly} through the small scales $\omega$ or $\gamma H_0$, whereas under NSOT it couples \textit{directly and strongly} through the large exchange scale $\gamma H_J$. This analysis of the effective dynamics of the Néel vector provides the physical reason behind the calculated gigantic response in the main text.

\end{document}